\documentclass[runningheads]{llncs}
\usepackage[T1]{fontenc}
\usepackage{graphicx}
\newcommand{\orcidMartin} {\orcidID{0000-0002-3696-9222}}
\newcommand{\orcidRaik}   {\orcidID{0009-0007-8348-6163}}
\newcommand{\orcidPatrick}{\orcidID{0000-0001-8178-4692}}

\usepackage[utf8]{inputenc}
\usepackage[T1]{fontenc}
\usepackage{slantsc}

\usepackage{graphicx}
\usepackage{amsmath}
\usepackage{amssymb}
\usepackage{stmaryrd}
\usepackage{xspace}
\usepackage{wrapfig}

\usepackage{xcolor}
\usepackage[hidelinks]{hyperref}

\usepackage{listings}
\usepackage{float}

\xdefinecolor{maincolor}{RGB}{0, 120, 140}
\colorlet{alertedcolor}{purple}
\colorlet{examplecolor}{green!50!black}

\usepackage{paralist}

\lstdefinestyle{pseudo}{language={},
  basicstyle=\normalfont,
  morecomment=[l]{//},
  morecomment=[s]{/*}{*/},
  morekeywords={for,to,while,do,if,then,else,each,end,Input,Output},
  mathescape=true,
  columns=fullflexible
}
\usepackage{booktabs}

\usepackage{cleveref}

\usepackage{array}
\usepackage[ruled,vlined,linesnumbered]{algorithm2e}
\usepackage{arydshln}
\usepackage{subcaption}
\usepackage{makecell}

\usepackage{tikz}
\usepackage{tkz-base}
\usepackage{tkz-euclide}
\usetikzlibrary{shapes.misc,calc,decorations,decorations.pathreplacing,backgrounds,shapes.multipart,fit,positioning}

\usetikzlibrary{arrows.meta,plotmarks}
\tikzset{
	event/.style={
		draw,
		inner sep=.5pt,
		circle,
		minimum width=6pt
	},
	events/.style={
		font=\tiny,
		xscale=.5,
		yscale=.3
	},
	unit/.style={
		event,
		path picture={ 
			\draw[black]
			(path picture bounding box.south east) -- (path picture bounding box.north west)
			(path picture bounding box.south west) -- (path picture bounding box.north east);
		}
	}
}

\usetikzlibrary{%
  bending,%
  decorations.markings,%
  decorations.pathmorphing}

\makeatletter
\newcounter{streamdiagram@y}
\newenvironment{streamdiagram}[1][]{
  \tikzpicture[streamdiagram,every legend/.style={},#1]
  \setcounter{streamdiagram@y}{0}
  \newcommand{\y}{-\value{streamdiagram@y}}
  \newcommand{\down}[1][1]{\addtocounter{streamdiagram@y}{##1}\renewcommand{\y}{-\value{streamdiagram@y}}}
  \newcommand{\x}[1]{(##1,\y)}
  \newcommand{\legend}[2][]{\down\node[left,every legend,##1] at \x0 {##2};}
  
}{
  \endtikzpicture
}
\makeatother
\pgfdeclarelayer{background}
\pgfsetlayers{background,main}
\pgfkeys{%
  /tikz/on layer/.code={
    \pgfonlayer{#1}\begingroup
    \aftergroup\endpgfonlayer
    \aftergroup\endgroup
  },
}
\tikzset{
  streamdiagram/.style={
    every label/.style={
      inner sep=2pt,
      line width=0pt
    },
    event/.style={
      circle,
      minimum height=10pt,
      minimum width=10pt,
      inner sep=-2pt,
      fill=##1
    },
    event/.default=red!20,
    caption/.style={
      append after command={
        (\tikzlastnode.center) node[font=\scriptsize] {##1}
      }
    }
  }
}

\newfloat{lstfloat}{htbp}{lop}
\floatname{lstfloat}{Listing}

\crefalias{lstfloat}{listing}
\DeclareCaptionSubType{lstfloat}
\newenvironment{proof*}[1]{\renewcommand{\proofname}{#1}\begin{proof}}{\end{proof}}
\newenvironment{proofof}[1]{\begin{proof*}{Proof of #1}}{\qed\end{proof*}}

\newcommand{\KWD}[1]{\ensuremath{\textrm{\textit{#1}}}}

\newcommand{\LolaName}{\textsc{Lola}}
\newcommand{\Lola}{\LolaName\xspace}

\newcommand{\TRUE}{\KWD{true}}
\newcommand{\FALSE}{\KWD{false}}

\def\orcidID#1{\smash{\href{http://orcid.org/#1}{\protect\raisebox{-1.25pt}{\protect\includegraphics{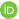}}}}}

\newcommand{\Bool}{\mathbb{B}}
\newcommand{\Real}{\mathbb{R}}
\newcommand{\Nat}{\mathbb{N}}
\newcommand{\Int}{\mathbb{Z}}

\newcommand{\Time}{\mathbb{T}}
\newcommand{\DD}{\mathbb{D}}
\newcommand{\seq}[1]{\langle #1\rangle}

\newcommand{\EE}{\mathbb{E}}
\newcommand{\Expr}{\KWD{Expr}}
\newcommand{\syminst}[1]{[#1]_{\mathit{sym}}}

\newcommand{\constraintsT}[1]{[\varphi]^{\leq #1}_\Psi}
\newcommand{\constraints}{\constraintsT{t}}
\newcommand{\assump}{\alpha}

\newcommand{\comps}{\mcomps}

\newcommand{\lolaDPI}{(\varphi,\comps,\Psi)}
\newcommand{\diagnoses}{\mathbf{D}}
\newcommand{\diags}{\diagnoses}

\newcommand{\minDiags}{\diags_\text{min}}
\newcommand{\allDiags}{\diags_\text{all}}

\newcommand{\instDiag}[1]{$#1$-instant diagnosis\xspace}
\newcommand{\instDiags}[1]{$#1$-instant diagnoses\xspace}

\newcommand{\temporalDiag}[1]{$#1$-temporal diagnosis\xspace}
\newcommand{\temporalDiags}[1]{$#1$-temporal diagnoses\xspace}

\newcommand{\aggr}{\KWD{aggr}}

\newcommand{\tsFree}{\KWD{tsFree}}

\newcommand{\OK}{\KWD{ok}}
\newcommand{\LOAD}{\KWD{ld}}
\newcommand{\ACC}{\KWD{acc}}

\newcommand{\out}{\KWD{out}}

\newcommand{\kpastK}[1]{t{\downarrow} #1}
\newcommand{\kpast}{\kpastK{k}}
\newcommand{\ExOneSymbolic}{
\begin{streamdiagram}[yscale=.45,xscale=1]
  \legend{}
  \draw[|->, gray] \x0 node[left] {$\LOAD$} -- \x5;
  \node[event=blue!20, caption={$\LOAD^0$}] at \x1 {};
  \node[event=blue!20, caption=3] at \x2 {};
  \node[event=blue!20, caption=4] at \x3 {};
  \node[event=blue!20, caption=3] at \x4 {};

  \legend{}
  \draw[|->, gray] \x0 node[left] {$\ACC$} -- \x5;
  \node[event=blue!20, caption=$\LOAD^0$] at \x1 {};
  \node[event=blue!20, caption=$\LOAD^0\!\!+\!\!3$] at \x2 {};
  \node[event=blue!20, caption=$\LOAD{}^0\!\!+\!\!7$] at \x3 {};
  \node[event=blue!20, caption=10] at \x4 {};
  
  \legend{}
  \draw[|->, gray] \x0 node[left] {$\OK$} -- \x5;
  \node[event=green!20, caption=$\TRUE$] at \x1 {};
  \node[event=green!20, caption=$\TRUE$] at \x2 {};
  \node[event=orange!20, caption=$?$] at \x3 {};
  \node[event=red!20, caption=$\FALSE$] at \x4 {};
  
\end{streamdiagram}
}

\newcommand{\DiagnosisTrace}{
\begin{streamdiagram}[yscale=.45,xscale=1]
  \legend{}
  \draw[|->, gray] \x0 node[left] {$i_1$} -- \x4;
  \node[event=blue!20, caption={$[2,3]$}] at \x1 {};
  \node[event=blue!20, caption=$5$] at \x2 {};
  \node[event=blue!20, caption=$?$] at \x3 {};

  \legend{}
  \draw[|->, gray] \x0 node[left] {$i_2$} -- \x4;
  \node[event=blue!20, caption={$[2,4]$}] at \x1 {};
  \node[event=blue!20, caption=$7$] at \x2 {};
  \node[event=blue!20, caption=$?$] at \x3 {};

  \legend{}
  \draw[|->, gray] \x0 node[left] {$\out_D$} -- \x4;
  \node[event=blue!20, caption=$?$] at \x1 {};
  \node[event=blue!20, caption=$?$] at \x2 {};
  \node[event=blue!20, caption=$?$] at \x3 {};

  \legend{}
  \draw[|->, gray] \x0 node[left] {$\out_T$} -- \x4;
  \node[event=blue!20, caption=$?$] at \x1 {};
  \node[event=blue!20, caption=$?$] at \x2 {};
  \node[event=blue!20, caption=$?$] at \x3 {};

  \legend{}
  \draw[|->, gray] \x0 node[left] {$\out_{A_1}$} -- \x4;
  \node[event=green!20, caption=$\TRUE$] at \x1 {};
  \node[event=red!20, caption=$\FALSE$] at \x2 {};
  \node[event=orange!20, caption=$?$] at \x3 {};
  
  \legend{}
  \draw[|->, gray] \x0 node[left] {$\out_{A_2}$} -- \x4;
  \node[event=green!20, caption=$\TRUE$] at \x1 {};
  \node[event=green!20, caption=$\TRUE$] at \x2 {};
  \node[event=orange!20, caption=$?$] at \x3 {};
  
\end{streamdiagram}
}

\usepackage{latexsym,footnote}
\usepackage{mathnotation}
\usepackage{array}
\usepackage{dashrule}
\usepackage{multirow}
\usepackage{tabularx}
\usepackage{cuted}
\usepackage{enumerate}
\usepackage{enumitem}
\usepackage{color}
\usepackage{arydshln}
\usepackage{cancel}
\usepackage{rotating}
\usepackage{framed}
\usepackage{ marvosym }
\usepackage{url}
\usepackage{amsfonts}
\usepackage{wasysym}
\usepackage{environ}

\usepackage{wrapfig}

\usepackage{comment}

\newcommand{\mD}{{\bf{D}}}

\newcommand{\dg}{\ensuremath{\Delta}\xspace}

\newcommand{\sd}{\textsc{sd}\xspace}
\newcommand{\mcomps}{\textsc{comps}\xspace}

\newcommand{\obs}{\textsc{obs}\xspace}
\newcommand{\dpi}{\mathsf{DPI}\xspace}

\newcommand{\ab}{\textsc{ab}}

\newcommand{\sdaa}[1]{\textsc{sd}^*[#1]}

{}
{}

\begin{document}

\title{A Unified Framework for Runtime Verification and Model-Based Diagnosis in \Lola}
%
\titlerunning{Runtime Verification and Model-Based Diagnosis in \Lola}
%

\author{
   Raik Hipler\inst{1}\orcidRaik \and
   Martin Leucker\inst{1}\orcidMartin \and 
   Patrick Rodler\inst{2}\orcidPatrick
}
\institute{
  University of Luebeck, Luebeck, Germany\\
  \email{\{hipler,leucker\}@isp.uni-luebeck.de}
  \and
  University of Klagenfurt, Klagenfurt, Austria\\
  \email{patrick.rodler@aau.at}
}
\authorrunning{R. Hipler et al.}
\maketitle
\begin{abstract}

Runtime verification has proven effective for detecting failures, while model-based diagnosis is a powerful method for identifying their underlying causes. 
Combining these two techniques is therefore highly valuable, as it can ultimately support mitigation measures for detected failures.
This paper proposes an integrated approach that specifies both runtime verification and diagnostic information within the well-known stream-based language \Lola. 
The resulting unified framework offers the expression of various
verification and diagnosis scenarios, as well as the combination of symbolic, stream-based reasoning under uncertainty with diagnostic inference for automated fault localization.

%

\keywords{Runtime Verification \and \Lola\ \and Uncertainties \and Assumptions \and Model-Based Diagnosis \and Multiple Observations \and Temporal Diagnosis}
\end{abstract}


\section{Introduction}
\label{sec:intro}

Beyond the correctness of safety-critical systems, their resilience is of paramount importance. In particular, systems must react appropriately to faults occurring during operation. To this end, failures must first be \emph{detected}, i.e., the system must recognize that something is going wrong, and then \emph{localized}, meaning that the root cause of the malfunction (e.g., a faulty component) must be determined.

\emph{Runtime verification (RV)} \cite{leucker09brief} is actively researched 
and well suited for detecting deviations from the specified behavior of a system during runtime. Similarly, \emph{model-based diagnosis (MBD)} \cite{Reiter87} is a well-founded, principled framework for localizing faults when a system no longer behaves as intended. Integrating RV with MBD promises a synergistic combination:  
automated fault detection and systematic fault localization in a live system monitoring framework.

Early work by \cite{BLS06} combined LTL-based RV monitors with MBD, using detected violations to compute minimal sets of abnormal components. However, monitoring and diagnosis remain largely separate: 
RV analyzes system behavior over potentially unbounded state sequences, whereas diagnosis is performed at a single, unspecified point in time after faults have been detected.

Köhl et al.~\cite{DBLP:journals/tecs/KohlH23} propose event-based diagnosis using a single model for both normative and failure actions, enabling simultaneous monitoring and diagnosis. 
While this identifies which failure actions occurred, it does not directly map them to abnormal components in the MBD sense.

A prominent line of work in RV is \emph{stream-based RV}, modeling monitoring as synchronous transformations of input streams to output streams via equational specifications. This paradigm, pioneered in \Lola~\cite{DBLP:conf/time/DAngeloSSRFSMM05}, has recently been extended to handle uncertain inputs and environmental assumptions, allowing refined inference based on expert knowledge \cite{HIPLER2026108004}. Observations are unfolded on the fly and evaluated using an SMT solver to produce verdicts at each time point.

In this paper, we integrate stream-based RV with MBD based on \Lola, yielding a sound, unified framework for online diagnosis. Our contributions are:

\begin{enumerate}[noitemsep,topsep=0pt]
	\item \textbf{Unified Encoding.} System descriptions 
	are expressed as specialized 
	streams which 
	implicitly represent all diagnoses for observed system behavior. Constraints generated and maintained in 
	\Lola are combined with an SMT solver and standard diagnosis algorithms to derive minimal diagnoses at runtime.
	\item \textbf{Multi-Instant Diagnosis.} 
	To explain time-invariant faults, 
	e.g., when the observation horizon is negligible compared to typical component lifetimes,
	we propose a theory dubbed \emph{multi-instant diagnosis}, which subsumes the well-known \emph{diagnosis over multiple observations}~\cite{Ignatiev_Morgado_Weissenbacher_Marques-Silva_2019} as a special case. 
	\item \textbf{Temporal Diagnosis.} By accounting for time-varying abnormality, our framework supports \emph{temporal diagnosis} \cite{friedrich_diagnosing_1991}, 
	identifying not only faulty components but also when faults occur. This is critical for systems with transient failures, e.g., overheated components or intermittently failing network links.
	\item \textbf{Various Other Diagnostic Scenarios.} We discuss further diagnostic settings that can be elegantly addressed in the \Lola RV framework---e.g., background knowledge, abstract modeling, as well as fault models and types.
	\item \textbf{Implementation and Evaluation.} We prototypically implemented the framework and validated it on two circuits from \mbox{ISCAS85}~\cite{ISCAS85} and ISCAS89~\cite{ISCAS89} benchmarks, 
	showing its feasibility 
	and providing first performance metrics.
\end{enumerate}
Overall, our work provides a compact, well-founded integration of RV and MBD, supporting a multitude of diagnostic use cases in a single, stream-based framework, bridging theoretical rigor with practical runtime applicability.


\section{Preliminaries}
\label{sec:prelims}

\subsection{Model-Based Diagnosis (MBD)}

\label{subsec:MBD}

We briefly review the general MBD framework \cite{Reiter87}, based on \cite{rodler_reducing_2017}. 
Let $\mathcal{L}$ be a decidable monotonic knowledge representation language. Throughout this paper, all formulas are assumed to be over some fixed $\mathcal{L}$.

\noindent\textbf{Diagnosed System.} 
The subject of a diagnosis task is a \emph{system}, e.g., a digital circuit or physical device.
%
	Formally, a system is a tuple
	$(\sd, \mcomps)$ where $\sd$, the system description, is a set of formulas, 
	and  $\mcomps$, the system components, is a finite set of constants. 	
%
The distinguished 
``abnormal'' predicate $\ab$ is used in $\sd$ to model the expected behavior $beh_c$ of components $c \in \mcomps$. 
So, 
$\sd$
can be split into two partitions, one, $\sd_{beh}$, that includes all formulas 
in which the $\ab$ predicate appears 
(\emph{behavioral knowledge}), and the other, $\sd_{gen}$, 
including all remaining formulas (\emph{general knowledge}).
Let, for now, $\sd_{beh} := \{\lnot\ab(c) \to beh_c $ $\mid c \in \mcomps\}$, which  
subsumes one statement of the form ``if $c$ is normal, 
then its  
\setlength{\intextsep}{0.45\baselineskip}
\begin{wrapfigure}{r}{0.47\linewidth}
	\centering
	\includegraphics[width=\linewidth]{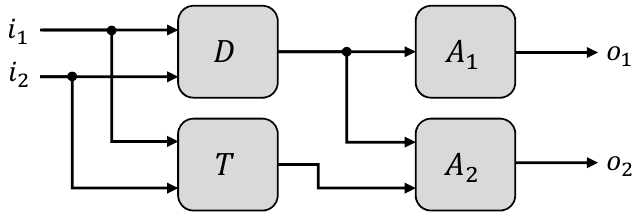}
	\captionsetup{skip=3pt}
	\caption{Running Example}
	\label{fig:ex}
\end{wrapfigure}
behavior is $beh_c$'' for each 
$c \in \mcomps$. 
In contrast, $\sd_{gen}$ can comprise,
e.g., formulas describing the system domain or the interplay between components.


\begin{example}
	\label{ex:running:system_description}
        \label{ex:running}
        Consider an alarm system for a medicine cooling room equipped
        with two cooling units and two temperature sensors, each
        located next to a cooling unit. To identify potential failures
        of the cooling 
        or the sensors, the system with
        components $\mcomps =$ $\{D, T, A_1, A_2\}$ shown
        in Fig.~\ref{fig:ex} is used. The system inputs are the
        numerical readings $i_1, i_2$ from the two temperature
        sensors.  
        $D$ computes the deviation between the
        sensor readings, and $A_1$ raises an alarm if this deviation
        reaches a safety threshold $d_{\max}$ (e.g., indicating a
        malfunction of one cooling unit and prompting
        inspection). 
        $T$ computes the average room
        temperature, and $A_2$ raises an alarm if both sensors agree
        on a critical temperature of at least $h_{\max}$ (e.g.,
        suggesting failure of both cooling units or an improperly
        closed door).  Formally, $beh_D := \out(D) = |i_1 - i_2|$,
        $beh_T := \out(T) = (i_1 + i_2) / 2$,
        $beh_{A_1} := \out(A_1) = (\out(D) \geq d_{\max})$, and
        $beh_{A_2} := \out(A_2) = (\out(T) - \out(D) \geq h_{\max})$.
        Together, these define $\sd_{beh}$. 
The system outputs are 
$o_1 = \out(A_1)$ and $o_2 = \out(A_2)$, which are part of $\sd_{gen}$.
\end{example}

\noindent\textbf{Diagnosis Problem.} The system description can model different behaviors depending on the components' states. Let
$
\sdaa{\dg} := \sd \cup \setof{\ab(c) \mid c \in \dg} \cup \setof{\lnot \ab(c) \mid c \in \mcomps \setminus \dg}
$
for $\dg \subseteq \mcomps$ denote the behavior description of the system $(\sd,\mcomps)$ under the assumption that components in $\dg$ are abnormal and all others are normal. The \emph{expected behavior} of a correctly working system is thus given by $\sdaa{\emptyset}$.
The core idea of MBD is to contrast this expected behavior with
\emph{observations} of the actual system behavior, represented by a
finite set of formulas $\obs$ (e.g., 
encoding inputs and outputs). 
A system together with observations constitutes a \emph{diagnosis problem instance (DPI)} $(\sd,\mcomps,\obs)$.

\noindent\textbf{Fault Detection and Localization.} Testing whether the expected behavior contradicts the observations, formally $\sdaa{\emptyset} \cup \obs$ being inconsistent, is called \emph{fault detection}. 
If positive, the goal of \emph{fault localization} is to identify the components causing 
the misbehavior.
To this end, MBD employs the notion of a \emph{diagnosis}, i.e., an assumption about the state (normal or abnormal) 
of each component that would explain the discrepancy between expected and observed behavior.
Formally, 
$\dg\subseteq\mcomps$ is a diagnosis for DPI $(\sd,\mcomps,\obs)$ iff $\sdaa{\dg} \cup \obs$ is consistent; it is \emph{minimal} iff there is no diagnosis $\dg'\subset\dg$.

Given $\sd_{beh}$ as defined above, only the normal behavior of each component is specified.
This is 
referred to as \emph{weak-fault model (WFM)}~\cite{rodler_how_should_2023}, and our assumption throughout this work. Using a WFM, minimal diagnoses are representative of all diagnoses, i.e., a set is a diagnosis iff it is a superset of some minimal diagnosis.
Hence, fault localization in this case focuses only on minimal diagnoses.

\noindent\textbf{Diagnosis Computation.} 
Basically, the size of the search space for minimal diagnoses is in $\mathcal{O}(2^{|\mcomps|})$.
A key concept to restrict this search space is that of a \emph{conflict} \cite{Reiter87,dekleer1987}, i.e., a set of components that cannot all be normal given $\obs$.
Formally, $C\subseteq\comps$ is a \emph{conflict} for a DPI $(\sd,\mcomps,\obs)$ iff $\sd \cup \obs \cup \setof{\lnot\ab(c) \mid c \in C}$ is inconsistent; it is \emph{minimal} iff there is no conflict $C'\subset C$.

Clearly, any explanation of the system misbehavior must assume at least one component in each minimal conflict to be abnormal. In other words \cite{Reiter87}: \emph{Every (minimal) diagnosis for a DPI is a (minimal) hitting set of all minimal conflicts for this DPI}. A \emph{(minimal) hitting set} of a collection $S=\setof{S_1,\dots,S_n}$ is a ($\subseteq$-minimal) set $H$ such that $H \subseteq \bigcup_{S_i \in S} S_i$ and $H \cap S_i \neq \emptyset$ for all $i=1,\dots,n$. Hence, a common way of diagnosis computation is  by first identifying minimal conflicts (e.g., using \cite{junker04,rodler_formal_2022}) and then calculating their minimal hitting sets.

Various diagnosis computation techniques are available for fault localization, including hitting-set--based best-first approaches \cite{Reiter87,greiner1989correction,rodler_memory-limited_2022,rodler_dynamichs_2023}, knowledge compilation methods \cite{Darwiche2001,Torasso_Torta_2006,Metodi_Stern_Kalech_Codish_2014}, stochastic strategies \cite{feldman_computing_2008,Rodler_Teppan_Jannach_2021}, and direct techniques \cite{Shchekotykhin2014,rodler_random_2022}. A comprehensive overview is provided in \cite{rodler_how_should_2023}.

\begin{example}
	\label{ex:runningDiagnosis}
	Recall the alarm system from Ex.~\ref{ex:running:system_description}, and let $d_{\max} := 1$ and $h_{\max} := 5$. Assume we observe its inputs and outputs as $(i_1,i_2,o_1,o_2) = (5,7,\false,\true)$. 
	First, a \emph{fault detection} is executed: Assuming all components to be normal, the predicted system behavior is i.a.\ 
	$out(D) = 2$ (using $i_1,i_2$) and $o_1 = out(A_1) = \true$ (using $out(D)$), 
	which contradicts the observation $o_1 = \false$. Hence, there must be some fault in the system.
	For \emph{fault localization}, conflict computation first yields two \emph{minimal conflicts} $C_1 = \{D,A_1\}$ (see derivation above) and $C_2 = \{D,T,A_2\}$ (because assuming these three components to be normal, $o_2 = \false$ can be inferred). Second, the computation of minimal hitting sets of $\{C_1,C_2\}$ yields the \emph{minimal diagnoses} $\dg_1 = \{ D \}$, $\dg_2 = \{ A_1,T \}$, and $\dg_3 = \{ A_1,A_2 \}$.    
\end{example}

\noindent\textbf{Multiple Diagnoses.} In practice, many minimal diagnoses may exist, and computing them is NP-hard \cite{Bylander1991}. As distinct diagnoses encode mutually exclusive assumptions, the goal is to identify the \emph{actual diagnosis}, i.e., the faulty components, and dismiss \emph{spurious} ones. This is addressed via \emph{ranking} (ordering diagnoses, e.g., by minimal cardinality or maximal probability), \emph{filtering} (computing only top-ranked diagnoses), and \emph{discrimination} (acquiring observations to prune the diagnosis space). The latter defines the \emph{sequential diagnosis problem} \cite{dekleer1987,rodler_sequential_2023-1}: Given $\dpi=(\sd,\mcomps,\obs)$, find observations $\obs_{\mathit{new}}$ such that a single (highly probable) minimal diagnosis remains for $(\sd,\mcomps,\obs\cup\obs_{\mathit{new}})$.

\noindent\textbf{Multiple Observations.} 
In the classic MBD setting, one considers a single observation at one time point. In scenarios such as RV, however, the system can be observed multiple times 
(possibly under different inputs or revealing previously unobserved 
values). Depending on whether component (ab)normality is assumed \emph{static} or \emph{dynamic} across observations, this yields a \emph{multi-observation (MO) diagnosis} or a \emph{temporal diagnosis} setting. The MO setting is relevant in systems like circuits, where component lifespans far exceed 
the observation period, while the temporal setting is useful for problems with frequent faults or long observation periods, e.g., networks with intermittently failing routers.

In MO diagnosis, given $\obs_1,\dots,\obs_k$, the goal is to find assumptions about component (ab)normality consistent with \emph{all} observations. Formally, $\Delta \subseteq \mcomps$ is a \emph{(minimal) MO-diagnosis} iff $\sdaa{\Delta} \cup \obs_i$ is consistent for all $i=1,\dots,k$ (and no strict subset satisfies this). Existing methods \cite{Kalech_Stern_Lazebnik_2021,Ignatiev_Morgado_Weissenbacher_Marques-Silva_2019} either encode and solve directly, or solve the $k$ single-observation DPIs and merge results. The latter uses \emph{minimal aggregated diagnoses}, which coincide with minimal MO-diagnoses: $\Delta\subseteq\mcomps$ is such a diagnosis \cite{Ignatiev_Morgado_Weissenbacher_Marques-Silva_2019} iff $\Delta=\bigcup_{i=1}^k \Delta_i$, where $\Delta_i\in\mD_i$ and $\mD_i$ is the set of minimal diagnoses for $\obs_i$, and no strict subset satisfies this.

In temporal diagnosis, component states may change over time. Given time-stamped observations $(\obs_1,t_1),\dots,(\obs_k,t_k)$, the goal is to find assumptions about component (ab)normality \emph{at each observation time point} consistent with all observations. Formally, a \emph{temporal diagnosis} \cite{friedrich_diagnosing_1991} 
is a set $\Delta_t \subset \mcomps \times \{t_1,\dots,t_k\}$, where $(c,t_i)\in\Delta_t$ denotes $\ab(c)$ at $t_i$ and $(c,t_i)\notin\Delta_t$ means $\lnot\ab(c)$ at $t_i$, such that $\sd \cup \bigcup_{i=1}^k (\obs_i,t_i) \cup \Delta_t$ is consistent.

\subsection{Symbolic Runtime Verification (RV) using \Lola}
\label{subsec:LOLA}

We briefly summarize symbolic RV with \Lola based on~\cite{HIPLER2026108004}.

\noindent\textbf{\Lola Syntax.}
A \emph{(synchronous) stream} is a function $s\colon\Time\to\DD$ assigning each instant in a time domain $\Time$ a value of a type $\DD$ (e.g., Boolean $\Bool$, real $\Real$), where $\Time$ is infinite ($\Nat$) or finite ($\{0,1,\dots,t_\KWD{max}\}$).
We write streams as sequences, e.g., $s\colon\{0,1,2\}\to\Real$ with $s(0)=4, s(1)=3, s(2)=2$ is written as $s=\seq{4,3,2}$.
%

A \emph{\Lola specification} transforms \emph{input streams} to defined 
streams~\cite{DBLP:conf/time/DAngeloSSRFSMM05}.
The values of \emph{defined streams} 
are obtained by evaluating expressions that depend on the values of (other) streams.
Furthermore, a special Boolean \emph{assumption stream} holds by definition, allowing background knowledge to be encoded and thereby enabling more precise reasoning. Finally, \emph{internal streams} are neither given as inputs nor defined by an expression, but implicitly defined via the assumption stream.
%
  Formally, a \Lola specification is a tuple $\varphi = (I,D,J,E,\assump)$ consisting of pairwise disjoint sets $I, D, J$ of typed identifiers for input streams, defined streams and internal streams, respectively; a distinguished assumption stream identifier $\assump$ of type $\Bool$;
  and a function $E$ assigning to every $y \in D\cup\{\assump\}$ of type $\DD$ a defining \emph{\Lola expression} $E_y$ of type $\DD$.   

  The \emph{set of all \Lola expressions} of type $\DD$ is denoted by $\EE_{\DD}$ and is recursively defined as
  $
  \EE_{\DD} := c \mid s[o|c] \mid f(e_1, \dots, e_n)
  $
  where 
  $c$ is a constant of type $\DD$, $s \in I\cup D\cup J$ is a stream identifier of type $\DD$, $o \in \Int$ is an offset, $f$ is a function symbol mapping from types $\DD_1, \dots, \DD_n$ to $\DD$, and $e_i\in\EE_{\DD_i}$.
The offset operator $s[o|c]$ represents the stream that has at time $t$ the value of stream $s$ at time $t+o$, and the value $c$ if $t+o\notin\Time$.
In case of $o=0$, we simply write $s$. 
A \Lola specification is called \emph{atemporal} iff every offset $o=0$.
An input stream can be \emph{certain}, i.e., a deterministic sequence of constants, or \emph{uncertain}, i.e., a (partially) nondeterministic sequence of values, e.g., due to noisy sensors, missing observations, or other sources of inaccuracy.
We call a (possibly nondeterministic) sequence of values for a set of streams over a sequence of time instants a \emph{trace}. 


\begin{figure}[t]
  \begin{minipage}[b]{.5\linewidth}
	\vspace{-1.5cm}
    \begin{lstlisting}[mathescape]
in  $\LOAD$ : $\Real$
def $\ACC := \ACC[-1|0]+\LOAD - \LOAD[-3|0]$
def $\OK \,\; := \ACC \leq 8$
$\assump:= 0\leq \LOAD\leq 5$
      \end{lstlisting}
    \captionof{lstfloat}{\label{lst:lolaex} Simple \Lola specification}
  \end{minipage}\hfill
  \begin{minipage}[b]{.5\linewidth}
    \ExOneSymbolic
    \caption{\label{fig:lolaex} Trace for Listing~\ref{lst:lolaex}}
  \end{minipage}
\end{figure}

\begin{example}
	\label{ex:lola_syntax}
	Assume a \Lola specification $\varphi=(\{\LOAD\},\{\ACC,\OK\},\emptyset,E,\assump)$ with $E_y$ for $y \in \{\ACC,\OK,\assump\}$ as defined in Listing~\ref{lst:lolaex}, and a corresponding trace shown in Fig.~\ref{fig:lolaex}.
	The defined stream $\ACC$ accumulates CPU 
	load measurements from the input stream $\LOAD$ by adding the current load to the previous accumulated value 
	and subtracting the third-last input value. Note, $\ACC[-1|0]$ refers to the value $\ACC$ attained one time instant earlier; it starts with 0 at $t=0$.
	The defined stream $\OK$ checks if the current $\ACC$ is at most 8.
	Due to $\assump$, $\LOAD$ is known to be always in $[0,5]$.
	Moreover, $\LOAD$ is an uncertain input stream as its first value is unknown.
\end{example}

\noindent\textbf{\Lola Semantics.} 
In case all input streams are certain, the \Lola semantics deterministically maps all input streams to unique defined streams~\cite{DBLP:conf/time/DAngeloSSRFSMM05}.
Given uncertain input streams, in contrast, there are generally multiple possible values input, defined, and internal streams can attain at each given time instant. To represent the possible values of a stream $s\in I\cup D\cup J$ at an instant $t\in\Time$, the concept of an \emph{instant variable} $s^t$ is used (cf.\ $\LOAD^0$ in Fig.~\ref{fig:lolaex}). 

Even though nondeterministic, values of uncertain input streams might nevertheless be constrained. To capture this, \Lola allows to specify \emph{input constraints}, (in)equalities over input stream instant variables specifying all possible input values at a given time. More specifically, given an input stream identifier $x\in I$, a constant $c$ can be represented as constraint $x^t=c$ (deterministic case), a closed interval $[l,r]$ as $l\leq x^t\leq r$ (analogously for open intervals), and a finite set $A$ as $\bigvee_{a\in A}x^t=a$. An entirely unknown input 
is simply represented by not specifying any input constraint for $x^t$. The \emph{set of all input constraints at time $t$} is denoted by $\psi^t$ (e.g., $\psi^0=\emptyset$ and $\psi^1=\{\LOAD^1=3\}$ in Fig.~\ref{fig:lolaex}).

%
%
%
Since input streams are represented by constrained instant variables instead of fixed values, the values of defined streams depend on these variables. This dependence is captured using \emph{symbolic expressions}, i.e., algebraic expressions over instant variables. For a type $\DD$, we write $\Expr_\DD$ for the set of all symbolic expressions of that type.
The symbolic expression for each defined stream $y \in D$ is characterized by its \Lola expression $E_y$. Hence, for 
each instant $t \in \Time$, this yields a constraint of the form $y^t = \syminst{E_y}(t)$. Here, the function $\syminst{e} \colon \Time \to \Expr_\DD$ returns the \emph{symbolic instantiation} of a \Lola expression $e \in \EE_\DD$ at time $t$ 
(e.g., for $t=2$ in Listing~\ref{lst:lolaex}, $X := \{\ACC^2=\ACC^1+\LOAD^2-0, \OK^2=\ACC^2\leq 8\}$ includes the constraints obtained in this way for $\ACC$ and $\OK$).
%
%
%
The symbolic instantiation is recursively defined as $\syminst{c}(t)=c$ for a constant $c$,  $\syminst{f(e_1,\dots,e_n)}(t)=f(\syminst{e_1}(t),\dots,\syminst{e_n}(t))$, $\syminst{s[o|c]}(t)=s^{t+o}$ if $t+o\in\Time$, and $\syminst{s[o|c]}(t)=c$ if $t+o\not\in\Time$. Similarly as for defined streams, the \Lola expression $E_\assump$ assigned to the assumption stream $\assump$ 
induces a constraint for every instant $t \in \Time$ by computing the symbolic instantiation $\syminst{E_\assump}(t)$ of $E_\assump$ (e.g., for instant $2$ in Listing~\ref{lst:lolaex}, $Y := \{0\leq\LOAD^2\leq 5\}$ includes the constraint derived from stream $\assump$). 

Putting together the constraint generation for defined streams and for the assumption stream, we obtain what is commonly referred to as the \emph{instantiation $\syminst{\varphi}(t)=\{y^t=\syminst{E_y}(t)\mid y\in D\}\cup\{\syminst{E_\assump}(t)\}$ of a \Lola specification} $\varphi=(I,D,J,E,\assump)$ for a time instant $t \in \Time$ (e.g., $\syminst{\varphi}(2)$ for the \Lola specification $\varphi$ given in Ex.~\ref{ex:lola_syntax} is equal to $X \cup Y$ for $X$ and $Y$ discussed above). 

Now, by combining the specification's instantiation with the input constraints over all time instants, formally $\bigcup_{t\in\Time} \syminst{\varphi}(t)\cup\psi^{t}$, all available information is captured. Based on this, the \Lola semantics specifies any combination of values of instant variables as a possible world (a \emph{model}) if it 
satisfies all these constraints, and as impossible otherwise (e.g., in Ex.~\ref{ex:lola_syntax}, the tuple of streams $(\langle1,3,4,3\rangle,\langle 1,7,$  $11,10 \rangle, \seq{\TRUE,\TRUE,\TRUE,\FALSE})$ for time instants $0\leq t\leq3$ corresponding to the stream identifiers $(\LOAD,\ACC,\OK)$ represents a model). We then say that some symbolic expression $e \in \Expr_\DD$ is entailed, formally $\bigcup_{t\in\Time} \syminst{\varphi}(t)\cup\psi^{t} \models e$, iff $e$ is true in all models (e.g., $\ACC^3 = 10$ 
is an entailment in Ex.~\ref{ex:lola_syntax}).  

\begin{example}
\label{ex:lola_semantics}
Recall the \Lola specification $\varphi$ from Ex.~\ref{ex:lola_syntax}. 
Despite the entirely unknown value of $\LOAD$ at instant $t=0$ (no input constraint for $\LOAD^0$), the constraint set $\bigcup_{t\in\Time} \syminst{\varphi}(t)\cup\psi^{t}$ allows to infer that $\LOAD^0\in[0,5]$ and thus $\OK^0$ and $\OK^1$ are $\true$. For $t=2$, however, $\ACC^2\in[7,12]$ is entailed, which is why $\OK^2$ is unknown, 
i.e., neither $\true$ nor $\false$ follows for $\OK^2$. At $t=3$, the constraint $\ACC^3 = \ACC^2 + \LOAD^3 - \LOAD^0$ involves a subtraction of (the only nondeterministic value) $\LOAD^0$. Hence, the deterministic values $\ACC^3 = 10$ and $\OK^3 = \false$ are implied.
\end{example}

\noindent\textbf{Symbolic Stream Reasoning in \Lola.}
In online monitoring, inputs are only known up to the current instant $t$; future inputs are unknown. 
In order to perform symbolic reasoning using the current information (as shown in Ex.~\ref{ex:lola_semantics}), the accumulated constraint set $\constraints := \bigcup_{t'=0}^t ( \syminst{\varphi}(t') \cup \psi^{t'} )$ 
needs to be stored and managed. 
A program that incrementally maintains and updates $\constraints$ is called a \emph{symbolic monitor} $M$, whose state at time $t$ is $M^t =\constraints$. When inputs for $t+1$ become available, $M^{t+1}$ is built by adding 
$\syminst{\varphi}(t+1)$ and $\psi^{t+1}$ to $M^{t}$. Using $M^t$, one can reason about the system up to time $t$ by means of an SMT solver (e.g., Z3~\cite{moura08Z3}), e.g., to determine the possible values of a stream.

Since new constraints are added at every instant, $M^t$ grows over time, 
leading to increasing runtime and memory consumption. 
In many applications, however, monitoring is required to run in 
\emph{constant} time and space, independent of $t$.
To achieve this, \emph{pruning strategies}~\cite{HIPLER2026108004} rewrite 
constraints so that the size of $M^t$ remains bounded. 
A symbolic monitor is \emph{sound} iff, after pruning, it allows to deduce at least the same values for a set of instant variables of interest as before pruning.
It is \emph{perfect} iff it deduces exactly the same values. 
A sound but imperfect monitor may overapproximate, i.e., allow additional values, which can be necessary to guarantee constant memory usage~\cite{HIPLER2026108004}.


\section{Combining RV and MBD using \Lola}
\label{sec:general}

In this section, our goal is to combine RV with MBD in the context of \Lola. This allows for a continuous observation of system executions, and for the reasoning about possible component faults, and thus synergizes the strengths of RV in symbolic stream-based reasoning and efficient fault detection with the merits of MBD as a powerful framework for fault localization.

In Sec.~\ref{subsec:systemAndSpec}, we expand on how and which diagnosis problems can be specified using \Lola. We then demonstrate system diagnosis in the stream-reasoning scenario. Sec.~\ref{subsec:single} considers component fault states assumed \emph{static} over the observation horizon and develops a generalized multi-observation diagnosis theory (cf.\ Sec.~\ref{subsec:MBD}). Sec.~\ref{subsec:temporal} focuses on \emph{dynamic} component fault states and the associated temporal diagnosis scenario in the RV context (cf.\ Sec.~\ref{subsec:MBD}). 
Sec.~\ref{sec:eval} demonstrates a proof-of-concept implementation.
Finally, Sec.~\ref{subsec:additional} briefly discusses additional diagnostic settings that can be elegantly represented in the \Lola framework—e.g., background knowledge, monitors, abstract system modeling, fault models, and different fault types with respect to their evolution over time. All proofs for the lemmas and theorems are included in Appendix~\ref{appendix}.

\subsection{Specifying Diagnosis Problems in \Lola}
\label{subsec:systemAndSpec}
%
In what follows, we consider diagnosed systems $S = (\sd,\comps)$ as per Sec.~\ref{subsec:MBD} which are \emph{synchronous} (all variables update simultaneously at each time step), \emph{discrete-time} (time progresses in steps, not continuously), and \emph{expressible in \Lola}.
Examples of system types that can be eligible are combinatorial and sequential circuits \cite{6771467}, spreadsheets \cite{rodler_choosing_2025-1}, software \cite{biere2009bounded},
(cyber)physical systems \cite{alur2015principles}, and discrete-event systems \cite{cassandras2007introduction}. 

To map $S$ to a \Lola specification $\varphi$, we have to decide how the system description  $\sd$ and the system components $\comps$ as well as the system variables (e.g., inputs, outputs, and internal values) can be captured by $\varphi$. Aligning with the syntax and semantics of \Lola expounded in Sec.~\ref{subsec:LOLA}, the basic idea is to
\begin{itemize}[noitemsep,topsep=0pt]
	\item specify $\sd$ 
	in terms of the assumption stream, which is true by definition,
	\item model all observable (but not necessarily observed) system aspects of interest 
	in terms of (uncertain) input streams, e.g., (noisy or intermittent) sensor information about system inputs, outputs or internal values, and
	\item define all unobservable system aspects we might be interested in inferring in terms of internal streams, e.g., the component (ab)normality states or internal system values not measurable or monitored by a sensor.
\end{itemize}
%
Defined streams, although not required to describe $S$, can be exploited, e.g., to specify auxiliary streams for 
system modeling,
or to define monitors \cite{leucker_brief_2009} that guard certain system aspects of interest, e.g., necessary invariants or integrity criteria. 
In this vein, if yielding a verdict $\false$, 
such monitors can serve as useful indicators of system malfunctions that trigger fault localization actions, as perpetual diagnostic reasoning might be infeasible due to system size or complexity.



\begin{definition}[\Lola System]
Let $S = (\sd,\mcomps)$ be a system as per Sec.~\ref{subsec:MBD} which is synchronous, discrete-time, and expressible in \Lola. Let $V_{o}(\sd)$ (and $V_{u}(\sd)$) be functions that extract all observable (unobservable) variables from the system description $\sd$.  
Then, the pair $(\varphi, \comps)$ is a \emph{\Lola system} for $S$ iff $\varphi=(I,D,J,E,\assump)$ is a \Lola specification where
\begin{compactitem}
	\item the assumption stream $\assump := \sd$ (\emph{system description}),
	\item there is an internal stream $c \in J$ of type $\Bool$ for each $c \in \comps$ where a value $\true$ indicates that component $c$ is abnormal (\emph{component health states}),
	\item there is an input stream $o \in I$ of type $\DD$ for each 
	$o \in V_o(\sd)$ of type $\DD$ (\emph{observable system variables}), and
	\item there is an internal stream $u \in J$ of type $\DD$ for each 
	$u \in V_u(\sd)$ of type $\DD$ (\emph{unobservable system variables}).
\end{compactitem}
\end{definition}

A \Lola system represents a mere description of the system to be diagnosed in terms of a \Lola specification. What is still missing are observations about the system. These are obtained by the input constraints  $\Psi = \bigcup_{t'=0}^{t} \psi^{t'}$ up to the current time instant $t$, because only these give (possibly uncertain) information about the actual input streams. Together with the \Lola system, this yields the full information needed to specify a \Lola diagnosis problem instance. Note that, in case of atemporal systems, where only 0-offsets
are used, all time instants can be considered separately and independently; i.e., only $\psi^{t}$, instead of the entire input constraint history, is relevant to the diagnostic reasoning at instant $t$.
\begin{definition}[\Lola DPI]\label{def:LOLA_dpi}
	Let $(\varphi, \comps)$ with $\varphi=(I,D,J,E,\assump)$ be a \Lola system and $\Psi=\bigcup_{t'=0}^t \psi^{t'}$ 
	be the set of input constraints for all streams $s \in I$ for the instants $0,\dots,t$.
	Then, the tuple $\lolaDPI$ is called a \emph{\Lola diagnosis problem instance (\Lola DPI)}.
\end{definition}

\begin{figure}[t]
\begin{minipage}[b]{.56\linewidth}
\begin{lstlisting}[mathescape]
internal $D,T,A_1,A_2$ : $\Bool$
in $i_1,i_2$ : $\Real$   $\quad$  in $\out_{A_1},\out_{A_2}$ : $\Bool$
internal $\out_D,\out_T$ : $\Real$
def $mon := \out_{A_2} \to \min(i_1,i_2) \geq 5$
$\assump := (\neg D\to (\out_D = |i_1 - i_2|))$
$\hspace{.45cm} \land\,(\neg T\to (\out_T = (i_1 + i_2)/2))$ 
$\hspace{.45cm} \land\,(\neg A_1\to (\out_{A_1} \leftrightarrow \out_D \geq 1))$ 
$\hspace{.45cm} \land\,(\neg A_2\to (\out_{A_2}\leftrightarrow \out_T-\out_D \geq 5))$ 
\end{lstlisting}
\captionof{lstfloat}{\label{lst:behSpec} \Lola specification}
\end{minipage}
\hfill
  \begin{minipage}[b]{.44\linewidth}
    \DiagnosisTrace
    \caption{\label{fig:diagTrace} Trace for Listings~\ref{lst:behSpec} and~\ref{lst:behSpecTemp}}
  \end{minipage}
\end{figure}

\begin{example}
\label{ex:specBeh}
Recall our alarm system from Fig.~\ref{fig:ex}, described in Examples~\ref{ex:running} and \ref{ex:runningDiagnosis}. Assume that the system's inputs ($i_1,i_2$) and outputs ($out_{A_1},out_{A_2}$) 
have attached (noisy) sensors and are thus observable, and let the internal variables ($out_{D},out_{T}$) be unobservable. Then, a \Lola DPI as per Def.~\ref{def:LOLA_dpi} corresponding to this system is given by $(\varphi,\comps,\Psi)$ where $\comps = \{D,T,A_1,A_2\}$, $\varphi$ is given in Listing~\ref{lst:behSpec}, and $\Psi = \{\psi^0,\psi^1,\psi^2\}$ is shown by the two top and the two bottom traces in Fig.~\ref{fig:diagTrace}. 
Note: 
\emph{(1)}~The sensors for the system inputs yield an output that is nondeterministic at instant $0$, deterministic at $1$, and entirely unknown at $2$, whereas the sensors readings for the output values are deterministic at $0$ and $1$, but entirely unknown at $2$ (cf.\ Fig.~\ref{fig:diagTrace}).
\emph{(2)}~The defined stream $mon$ 
is not part of the \Lola DPI, but constitutes a monitor of the integrity criterion that $\min(i_1,i_2) \geq 5$ must hold whenever $out_{A_2}$ is $\true$.
Thus, if the value of $mon$ is $\false$, \emph{fault detection} is positive, and \emph{fault localization} can be started, involving reasoning about possible sets of faulty components (values of the internal streams in $\comps$) 
using the symbolic monitor for $\varphi$ and $\Psi$ (cf.\ Sec.~\ref{subsec:LOLA}) along with an SMT solver and off-the-shelf MBD algorithms (cf.\ Sec.~\ref{subsec:MBD}).   
\end{example}

\subsection{Multi-Instant Diagnosis with \Lola}
\label{subsec:single}

Let us first consider the common setting where component states are static over time.
The basic idea underlying our diagnosis approach is that the symbolic monitor maintains a consistent constraint set $\constraints$ for current instant $t$ (see Sec.~\ref{subsec:LOLA}), from which diagnoses can be inferred.
Specifically, a diagnosis $\Delta$ encodes an assumption about each component's state, i.e., normal or abnormal, which is consistent with $\constraints$.
A component $c$ at instant $t'$ assumed to be normal ($c\not\in\Delta$) means that $\neg c^{t'}$ holds, and if assumed to be abnormal ($c\in\Delta$) $c^{t'}$ holds.
The question is for which instants $t'$ each component's state should be assumed.
Two natural options are to only consider the current instant ($t'=t$) or additionally also the entire past ($t'\leq t$).
The former does not make assumptions about past component behavior, while the latter assumes the same states throughout the past.
To generalize, we consider diagnoses conforming to $k$ past instants. 
We denote by $\kpast:=\{t'\in\Time\mid t-k\leq t'\leq t\}$ all instants from $t-k$ until $t$.

\begin{definition}[$k$-Instant Diagnosis]
  \label{def:lolaDiagnosisAtemp}
  Given a \Lola DPI $\lolaDPI$ and $k\in\Nat$, $\Delta\subseteq\comps$ is a \emph{\instDiag{k}} for $t\in\Time$ iff $\{c^{t'}\mid {c\in\Delta\land t'\in\kpast}\}\cup \{\neg c^{t'}\mid c\in\comps\setminus\Delta\land t'\in\kpast\}\cup\constraints$
  is consistent.
  A \instDiag{k} $\Delta$ is \emph{minimal} iff no $\Delta'\subset\Delta$ is a \instDiag{k}.
\end{definition}


Intuitively, \instDiag{0} is akin to diagnosis with a single observation (the current inputs), and \instDiag{k} with $k>0$ is similar to multi-observation diagnosis over $k$ observations~\cite{Ignatiev_Morgado_Weissenbacher_Marques-Silva_2019}.
In fact, as we will see later on, this connection holds up as long as $\varphi$ is atemporal, i.e., without non-zero offsets.

Now, an obvious question is why one should not always choose a large value for $k$.
The drawback is that a pruning strategy (cf.\ Sec.~\ref{subsec:LOLA}) must retain variables from the past $k$ instants; only variables older than $t-k$ can be pruned.
This increases the symbolic monitor's size.
Nevertheless, as long as $k$ is constant, a constant pruning strategy guarantees constant-time diagnosis, in the sense that complexity is independent of $t$.
Particularly, constant monitoring is impossible if the entire past up to $t$ ($k=t$) is considered.
However, when a fault is detected, it might be a sensible assumption that it is likely to have occurred rather recently.


\begin{table}[t]
  \centering
  \fontsize{8pt}{10pt}\selectfont
  \begin{tabular}{r||c|c|c}
                    & $t=0$                         & $t=1$                             & $t=2$ \\ \hline
    $0$-instant     & $\{D,A_1\}, \{T\},\{A_2\}$    & $\{D\},\{A_1,A_2\}, \{A_1,T\}$    & $\emptyset$    \\ \hline
    $1$-instant     & $\{D,A_1\}, \{T\},\{A_2\}$    & \makecell{$\{D,A_1\},\{D,T\},\{D,A_2\},$\\$\{A_1,T\},\{A_1,A_2\}$}    & $\{D\},\{A_1,A_2\}, \{A_1,T\}$    \\ \hline
    \makecell{$2$-instant/\\$\aggr$ of 0's}     & $\{D,A_1\}, \{T\},\{A_2\}$    & \makecell{$\{D,A_1\},\{D,T\},\{D,A_2\},$\\$\{A_1,T\},\{A_1,A_2\}$}   & \makecell{$\{D,A_1\},\{D,T\},\{D,A_2\},$\\$\{A_1,T\},\{A_1,A_2\}$}
  \end{tabular}
  \caption{
    \label{table:diagsAtemp} Minimal diagnoses for trace from Fig.~\ref{fig:diagTrace} on specification from Listing~\ref{lst:behSpec}.
    Row 3 also shows minimal aggregations of all \instDiags{0} for $0,\dotsc,t$.
  }
\end{table}

To get an intuition, 
let us first consider \instDiag{0} for atemporal $\varphi$.

\begin{example}
  \label{ex:single}
  Recall Ex.~\ref{ex:specBeh}. 
  We assume inputs for instants $t=0,1,2$ as given by Fig.~\ref{fig:diagTrace}.
  The minimal \instDiags{0} per $t$ are shown in Tab.~\ref{table:diagsAtemp}.
  
  At instant $t=0$, we have $\syminst{\varphi}(0) = \{(\neg D^0\to \out_D^0 = |i_1^0 - i_2^0|)\land (\neg T^0\to \out_T^0 = (i_1^0 + i_2^0)/2)\land\dotsc \}$.
  The sensors for the inputs deliver noisy values reflected by constraints $\psi^0=\{2\leq i_1^0\leq 3, 2\leq i_2^0\leq 4, \out_{A_1}^0=\TRUE, \out_{A_2}^0=\TRUE\}$.
  Thus, $\constraintsT{0}=\syminst{\varphi}(0)\cup\psi^0$.
  Since there is only evidence for $\out_{A_2}^0$ being incorrect ($\out_T^0-\out_D^0=[2,3.5]-[0,2]=[0,3.5] \not\geq 5$), the obvious conflict is $\{D,A_2,T\}$, leading to diagnoses $\{D\},\{A_2\},\{T\}$ (for $A_1$ we have $\out_D^0\in[0,2]$ which does not imply a fault).
  However, $\{D\}$ is actually not valid due to $D^0\land \neg T^0\land \neg A_1^0\land \neg A_2^0$ being inconsistent with $\constraintsT{0}$: $[2,3.5]-\out_D^0 \geq 5$ implies $\out_D^0$ being negative and $A_1$ therefore being abnormal, too (because the normal behavior of $D$ can only produce non-negatives). 
  
  For $t=1$, $\syminst{\varphi}(1)$ and $\psi^1=\{i_1^1=5, i_2^1=7, \out_{A_1}^1=\FALSE, \out_{A_2}^1=\TRUE\}$ are added, and diagnostic reasoning is identical to Ex.~\ref{ex:runningDiagnosis}.
  For $t=2$, since $i_1^2$ and $i_2^2$ are fully unknown, any assignment for the outputs $\out_{A_1}$ and $\out_{A_2}$ is consistent with $\constraintsT{2}$, leading to $\emptyset$ being the only minimal diagnosis.
\end{example}


Similar to how a multi-observation diagnosis is also a diagnosis for each single observation~\cite{Ignatiev_Morgado_Weissenbacher_Marques-Silva_2019}, since a \instDiag{k} makes assumptions about each component's state over the past $k$ instants, it is also a \instDiag{0} for the past $k$ instants. 
Particularly, if $k$ is set to the current instant $t$, it explains all observations for instants $0,\dotsc,t$.
Tab.~\ref{table:diagsAtemp} depicts the minimal \instDiags{k} for $k=1$ and $k=2$.
Since \instDiag{1} no longer reasons at $t=2$ about the component states for instant $0$, it cannot explain faults for that instant.


\begin{lemma}
  \label{lem:multiIsSingle}
  Let $\lolaDPI$ be a \Lola DPI. 
  Given $t\in\Time$ and $k\in\Nat$, every \instDiag{k} for $t$ is also a \instDiag{0} for every $t'\in\kpast$.
\end{lemma}

As mentioned in Sec.~\ref{subsec:MBD}, a common technique to obtain diagnoses for multiple observations is to compute the diagnoses for each observation separately and to aggregate them afterwards.
In fact, this also works with \instDiag{k} if $\varphi$ is atemporal (Theorem~\ref{thm:multiAggr} (ii)).
Let us denote by $\aggr(\diags_0,\dotsc,\diags_n)=\{\bigcup_{i=0}^{n} \Delta_i\mid \Delta_i \in \diags_i\}$ the set of aggregated diagnoses over sets of diagnoses $\diags_i$.
E.g., row 3 of Tab.~\ref{table:diagsAtemp} depicts for each $t$ the minimal aggregated \instDiags{0} for $0,\dotsc,t$;
they are identical to the \instDiags{2}.
However, if $\varphi$ involves non-zero offsets, i.e., temporal dependencies, this is not guaranteed:


\begin{lstfloat}[t]
\begin{lstlisting}[mathescape]
internal $D,T,A_1,A_2$ : $\Bool$               internal $\out_D,\out_T$ : $\Real$
in $i_1,i_2$ : $\Real$                     in $\out_{A_1},\out_{A_2}$ : $\Bool$
def $dev := |i_1 - i_2|$                       def $avg := (i_1 + i_2)/2$
$\assump := (\neg D\to (\out_D = \out_D[-1,0] + dev - dev[-3,0]))$
$\hspace{.45cm} \land\,(\neg T\to (\out_T = \out_T[-1,0] + avg - avg[-3,0]))$
$\hspace{.45cm} \land\,(\neg A_1\to (\out_{A_1} \leftrightarrow \out_D \geq 1))\land(\neg A_2\to (\out_{A_2}\leftrightarrow \out_T-\out_D \geq 5))$ 
\end{lstlisting}
\caption{Temporal extension of Listing~\ref{lst:behSpec}.}
\label{lst:behSpecTemp}
\end{lstfloat}

\begin{table}[t]
  \centering
  \resizebox{\columnwidth}{!}{
  \begin{tabular}{r||c|c|c}
                    & $t=0$                         & $t=1$                             & $t=2$ \\ \hline
    $0$-instant     & $\{D,A_1\}, \{T\},\{A_2\}$    & $\emptyset$    & $\emptyset$    \\ \hline
    $\aggr$ of 0's  & $\{D,A_1\}, \{T\},\{A_2\}$    & $\{D,A_1\}, \{T\},\{A_2\}$    & $\{D,A_1\}, \{T\},\{A_2\}$    \\ \hline
    $1$-instant     & $\{D,A_1\}, \{T\},\{A_2\}$    & \makecell{$\{D,A_1\},\{D,T\},\{D,A_2\},$\\$\{A_1,T\},\{A_1,A_2\}$}   & $\emptyset$\\ \hline
    $2$-instant     & $\{D,A_1\}, \{T\},\{A_2\}$    & \makecell{$\{D,A_1\},\{D,T\},\{D,A_2\},$\\$\{A_1,T\},\{A_1,A_2\}$}   & \makecell{$\{D,A_1\},\{D,T\},\{D,A_2\},$\\$\{A_1,T\},\{A_1,A_2\}$}\\ \hline
    $1$-temporal    & \makecell{$\{(D,0),(A_1,0)\},$\\$\{(T,0)\},\{(A_2,0)\}$} & \makecell{$\{(T,0),(A_1,1)\},\{(A_2,0),(A_1,1)\}$,\\$\{(D,0),(A_1,0)\},\{(T,0),(D,1)\},$\\$\{(A_2,0),(D,1)\}$} & $\emptyset$\\ \hline
    $2$-temporal    & \makecell{$\{(D,0),(A_1,0)\},$\\$\{(T,0)\},\{(A_2,0)\}$} & \makecell{$\{(T,0),(A_1,1)\},\{(A_2,0),(A_1,1)\}$,\\$\{(D,0),(A_1,0)\},\{(T,0),(D,1)\},$\\$\{(A_2,0),(D,1)\}$} & \makecell{$\{(T,0),(A_1,1)\},\{(A_2,0),(A_1,1)\}$,\\$\{(D,0),(A_1,0)\},\{(T,0),(D,1)\},$\\$\{(A_2,0),(D,1)\}$}
  \end{tabular}
  }
  \caption{
    \label{table:diagsTemp} Minimal diagnoses for trace from Fig.~\ref{fig:diagTrace} on specification from Listing~\ref{lst:behSpecTemp}.
    The second row shows minimal aggregations of all \instDiags{0} for $0,\dotsc,t$.
  }

\end{table}

\begin{example}
  \label{ex:multiTemp}
  Consider the diagnosis specification $\varphi$ from Listing~\ref{lst:behSpecTemp} which is an extension of Listing~\ref{lst:behSpec}, where $D$ is no longer specified to compute the deviation between the current $i_1$ and $i_2$ but to keep track of the sum of the three most recent deviations (analogously for $T$ and the average). 
  Assuming again inputs from Fig.~\ref{fig:diagTrace}, for $t=0$, reasoning remains unchanged compared to Ex.~\ref{ex:single}.
  However, for $t=1$, $\emptyset$ is the only minimal \instDiag{0} because no values for $D^0, T^0, A_1^0, A_2^0$ are assumed, i.e., the component states are undefined for the past.
  Specifically, $\out_D^1$ and $\out_T^1$ could in principle assume any value due to their dependence on the past and, as a result, no component has to be faulty at time 1.
  The minimal aggregated diagnoses up to $t=1$ are therefore identical to the \instDiags{0} for $t=0$ (see Tab.~\ref{table:diagsTemp}).
  However, among these, $\{T\}$ and $\{A_2\}$ are actually not \instDiags{1} for $t=1$ because assuming them for both $t=0$ and $t=1$ leads to inconsistency.
  E.g., if $\{T\}$ would be a diagnosis, ${\neg D^0\land \neg D^1\land \neg A_1^0\land \neg A_1^1}$ would need to be consistent with $\constraintsT{1}$.
  However, $\neg D^0\land\neg A_1^0$ implies $\out_D^0\in[1,2]$, hence $\neg D^1$ implies $\out_D^1\in[3,4]$, and finally $\out_{A_1}^1=\TRUE$ due to $\neg A_1^1$.
  This contradicts the observation that $\out_{A_1}^1=\FALSE$. 
  Therefore, either $D$ or $A_1$ must be abnormal together with $T$.
\end{example}

This example illustrates that an empty \instDiag{0} does not imply that there is no fault at all.
Instead, it only means that every potential fault could be explained by abnormal behavior in the past, i.e., there is no evidence for components to be faulty \emph{now}.
While in Ex.~\ref{ex:multiTemp} this leads to empty diagnoses, this is not necessarily the case and depends on the temporal interconnections between components and which of them are observed. 
As (i) in the following theorem shows, aggregating \instDiags{0} is, however, less ``focused'' than \instDiag{k} for $k>0$ if the specification is not atemporal.

\begin{theorem}
  \label{thm:multiAggr}
  Let $\lolaDPI$ be a \Lola DPI.
  Given $t\in\Time$ and $k\in\Nat$, let $\allDiags$ be the set of all \instDiags{k} and $\allDiags'=\aggr(\diags_{t-k},\dotsc,\diags_t)$ where $\diags_{t'}$ is the set of all \instDiags{0} for $t'\in\kpast$.
  Furthermore, let $\minDiags\subseteq\allDiags$ and $\minDiags'\subseteq\allDiags'$ be their respective subsets containing only minimal diagnoses.
  It holds that
  \begin{inparaenum}[(i)]
    \item $\allDiags\subseteq\allDiags'$,
    \item $\allDiags=\allDiags'$ if $\varphi$ is atemporal, and
    \item ${\forall\Delta\in\minDiags\colon\exists\Delta'\in\minDiags'\colon\Delta'\subseteq\Delta}$.
  \end{inparaenum}
\end{theorem}

Note that (i) of Theorem~\ref{thm:multiAggr} considers \emph{all} diagnoses, not only minimal ones:
A minimal \instDiag{k} might not be a minimal aggregated diagnosis due to not being a \instDiag{0} for all past $k$ instants.
However, according to (iii), for each minimal \instDiag{k} there exists a subset that is a minimal aggregation (compare rows 2 and 4 of Tab.~\ref{table:diagsTemp}).
Similarly, larger values for $k$ mean a diagnosis explains more instants at once (compare rows 1, 3, and 4 of Tab.~\ref{table:diagsTemp}):

\begin{theorem}
  \label{thm:multiPrecision}
  Let $\lolaDPI$ be a \Lola DPI.
  Given $t\in\Time$ and $k,k'\in\Nat$ with $k'\leq k\leq t$, let $\allDiags$ and $\allDiags'$ be the sets of all \instDiags{k} and all \instDiags{k'} for $t$, respectively.
  Furthermore, let $\minDiags\subseteq\allDiags$ and $\minDiags'\subseteq\allDiags'$ be their respective subsets containing only minimal diagnoses.
  It holds that
  \begin{inparaenum}[(i)]
    \item $\allDiags\subseteq\allDiags'$ and
    \item $\forall\Delta\in\minDiags\colon\exists\Delta'\in\minDiags'\colon\Delta'\subseteq\Delta$.
  \end{inparaenum}
\end{theorem}

\subsection{Temporal Diagnosis with \Lola}
\label{subsec:temporal}

While above approach allows to accurately diagnose the setting where component states are time-invariant,
often it might be that a component is only abnormal for \emph{some} time and recovers afterwards, e.g., due to overheating.
This allows faults caused by an abnormal component to manifest themselves at a later time when the component may behave correctly again.
In general, component states may change arbitrarily.
Therefore, we extend our approach by associating each component with the time of the fault:

\begin{definition}[$k$-Temporal Diagnosis]
  \label{def:lolaDiagnosisTemp}
  Given a \Lola DPI $\lolaDPI$ and $k\in\Nat$, a set $\Delta\subseteq\comps\times\kpast$ is a \emph{\temporalDiag{k}} for $t\in\Time$ iff
  $
    \{c^{t'}\mid (c,t')\in\Delta\}\cup \{\neg c^{t'}\mid (c,t')\in(\comps\times\kpast)\setminus\Delta\}\cup\constraints
  $
  is consistent.
  A \temporalDiag{k} $\Delta$ is minimal iff no $\Delta'\subset\Delta$ is a \temporalDiag{k}.
\end{definition}

\begin{example}
  \label{ex:temp}
  Consider again Listing~\ref{lst:behSpecTemp} with inputs from Fig.~\ref{fig:diagTrace}.
  Tab.~\ref{table:diagsTemp} depicts the minimal \temporalDiags{k} for $k=1,2$ and instants $t=0,1,2$.
  For all instants, the \temporalDiags{k} are, barring the timestamps, identical to the \instDiag{k} from before.
  However, using the additional time information, we can infer, e.g., from $\{(T,0),(A_1,1)\}$ that the observation can be explained when $T$ was previously abnormal and $A_1$ is abnormal now.
  Particularly, \temporalDiag{2} concludes with $\{(D,0),(A_1,0)\}$ at $t=1$ that the fault could be entirely in the past.
\end{example}

As the example shows, temporal diagnosis allows for more precise fault localization by pinpointing the possible times of a fault.
In fact, as (i) in the following theorem shows, for each \temporalDiag{k} there exists a $k$-instant timestamp-free counterpart, and vice versa.
We denote the omission of timestamps in temporal diagnoses $\diags$ as $\tsFree(\diags)=\{\{c\mid (c,t')\in\Delta\}\mid\Delta\in\diags\}$.

\begin{theorem}
  \label{thm:temp}
  Let $\lolaDPI$ be a \Lola DPI.
  Given $k\leq t$, let $\allDiags$ be the set of all \instDiags{k} and $\allDiags'$ the set of all \temporalDiags{k} for $t$.
  Furthermore, let $\minDiags\subseteq\allDiags$ and $\minDiags'\subseteq\allDiags'$ be their respective subsets containing only minimal diagnoses.
  It holds that
  \begin{inparaenum}[(i)]
    \item $\allDiags=\tsFree(\allDiags')$ and
    \item $\minDiags\subseteq\tsFree(\minDiags')$.
  \end{inparaenum}
\end{theorem}

Note that (i) in Theorem~\ref{thm:temp} considers \emph{all} diagnoses, not only minimal ones.
For minimal sets, (i) generally does not hold:
E.g., it is possible that $\minDiags'=\{\{(X,0)\},\{(Y,0),(X,1)\}\}$, i.e., $\tsFree(\minDiags')=\{\{X\},\{Y,X\}\}$, but $\minDiags=\{\{X\}\}$.
Here, the observation can be explained by the non-minimal temporal diagnosis $\{(X,0),(X,1)\}\supset\{(X,0)\}$, which is why $\{X\}$ is a \instDiag{k}.
While $(X,0)$ is sufficient and thus minimal, $(X,1)$ is not and must be accompanied by $(Y,0)$.
The \instDiag{k} cannot make this distinction.
But, as stated by (ii), every minimal \instDiag{k} is contained in the timestamp-free \temporalDiags{k}.

\subsection{Implementation and Evaluation}
\label{sec:eval}

We prototypically extended the symbolic \Lola tool used in \cite{HIPLER2026108004,HiplerKLS24} with the proposed diagnosis approaches as a proof of concept.
We implemented the standard HS-Tree algorithm~\cite{Reiter87,rodler_dynamichs_2023} together with \mbox{QuickXplain}~\cite{junker04,rodler_formal_2022} and Z3~\cite{moura08Z3} as backend solver.
The tool computes all minimal \instDiags{k} and \temporalDiags{k} per instant.
We evaluated our approach on two Boolean circuits, c17 and s27, from the ISCAS85~\cite{ISCAS85} and ISCAS89~\cite{ISCAS89} suites\footnote{
  The experiments were run on  a 64-bit Linux machine with an Intel Core i7-13650U CPU and 32 GB RAM.
  The tool and experiment data is available at \url{https://gitlab.isp.uni-luebeck.de/Raik.Hipler/lola-mbd-evaluation}.
}.
c17 has six components and is purely combinational (atemporal), whereas s27 has 14 components and is sequential. 
We translated both circuits into \Lola 
and injected stuck-at-$\FALSE$ faults for one component at a time, to produce erroneous outputs.
For c17, the fault is permanent;
for s27, the fault occurs at instant 3 and persists thereafter.

For c17, we diagnosed all 32 possible input combinations and their outputs for each injected fault in a single trace (since c17 is atemporal each instant is independent).
The actual diagnosis $\{c\}$ was identified as \instDiag{k} if the fault manifested in least one of the last $k$ instants.
Moreover, for faulty instants $t'$, $\{(c,t')\mid t'\in\kpast\}$ was always a \temporalDiag{k}.
%
For s27, where input order matters, we generated ten random traces of ten instants each to produce outputs for each injected fault.
Since faults may propagate over time, both approaches localize the actual fault only if its origin lies within the past $k$ instants. 
Because s27 is sequential and $c$ is faulty from instant 3 onwards, at time $t\geq 3$ localization of the actual diagnosis is only guaranteed for $k\geq t-3$.

As expected, larger values for $k$ increase runtimes.
For c17, \instDiag{k} required up to 246\,ms per instant for $k=0$ and 775\,ms for $k=31$;
for s27, up to 240\,ms for $k=0$ and 1106\,ms for $k=9$.
Since \temporalDiag{k} treats each $(c,t')$ as its own component, runtimes grow substantially:
for one injected fault in c17, up to 8.7\,s per instant were required for $k=5$, compared to up to 136\,s for $k=12$.
Because the circuits are Boolean, perfect pruning~\cite{HIPLER2026108004} retains complete information about instant variables $s^{t'}$ with $t'\in\kpast$.
With constant $k$ over the trace, runtimes remain independent of trace length.



\subsection{Extended Settings and Practical Considerations}
\label{subsec:additional}

Beyond the settings described here, our framework naturally captures additional scenarios, which we briefly outline and will study in future work:

\noindent\textbf{Background Knowledge.}
In MBD, a component $c$ known to be correct can be specified as background knowledge \cite{rodler_interactive_2015}, reducing the diagnostic search space. In a \Lola DPI, this is achieved by extending $\assump$ with $\land \lnot c$. 

\noindent\textbf{Fault Models.}
If $\sd_{beh}$ (cf.\ Sec.~\ref{subsec:MBD}) includes formulas 
$\ab(c)\rightarrow fbeh_{c,1}\lor\dots\lor fbeh_{c,k}$
where $fbeh_{c,i}$ are different faulty behaviors, $\sd$ is a \emph{strong fault model (SFM)} \cite[Chap.~10]{van_harmelen_handbook_2008}. This is naturally expressed in \Lola by extending $\assump$ accordingly. E.g., if component $T$ from Listing~\ref{lst:behSpec} is either stuck at $0$ or outputs $i_1$ if abnormal, then $T\to (\out_T=0)\lor(\out_T=i_1)$ is added to $\assump$. In this case, the different notion of a \emph{kernel diagnosis} \cite{DEKLEER1992197} (instead of minimal diagnosis) applies.

\noindent\textbf{Sequential Diagnosis.}
Discrimination between multiple diagnoses can be done by automatically defining informative \cite{rodler_active_2017,rodler_sequential_2023-1} monitors to prune the search space.

\noindent\textbf{Assumptions about Faults.}
Diagnostic efficiency can be improved by constraining the state evolution of a component $c$. \emph{Persistent faults}, which cannot recover automatically (e.g., a blown fuse), are specified by $c[-1|\FALSE]\to c$. \emph{Time-invariant faults} are specified by $c[-1|\FALSE]\to c \land c\to c[-1|\TRUE]$.

\noindent\textbf{Abstract Modeling.}
Besides precise value-based modeling, as discussed in this work, \Lola supports \emph{abstract dependency-based models} (cf.\ \cite{rodler_choosing_2025-1}), both instead or combined with exact models. In this vein, a user can trade a higher diagnostic efficiency for a (potentially) higher number of spurious diagnoses \cite{rodler_choosing_2025-1}.


\noindent\textbf{Diagnostic Timing.}
Due to the inherent complexity of MBD \cite{Bylander1991}, calculating diagnoses for every instant from can be practically hard or infeasible. Thus, a user can decide to diagnose only \emph{periodically}, \emph{on demand}, or \emph{after fault detection}.


\section{Conclusion}
\label{sec:conclusion}

We introduced a unified framework integrating MBD and stream-based RV using \Lola.
The approach enables online fault localization under possibly
uncertain observations by way of symbolic reasoning. 
A proof-of-concept implementation confirms feasibility.
%

\bibliographystyle{splncs04}
\bibliography{references,references_mbd}

\appendix
\section{Proofs for Section~\ref{sec:general}}
\label{appendix}

\begin{proofof}{Lemma~\ref{lem:multiIsSingle}}
  Let $\Delta$ be a \instDiag{k} for $t$.
  Furthermore, let $A:=\{c^{t'}\mid c\in\Delta\land t'\in\kpast\}$ and $B:=\{\neg c^{t'}\mid c\in\comps\setminus\Delta\land t'\in\kpast\}$.
  By Def.~\ref{def:lolaDiagnosisAtemp}, $A\cup B\cup\constraints$ is consistent.
  Since for every $t'\in\kpast$, we have $\{c^{t'}\mid c\in\Delta\}\subseteq A$ and
  $\{\neg c^{t'}\mid c\in\comps\setminus\Delta\}\subseteq B$,
  $\{c^{t'}\mid c\in\Delta\}\cup \{\neg c^{t'}\mid c\in\Delta\}\cup\constraintsT{t'}$ must be consistent as well, and therefore $\Delta$ a \instDiag{0} for $t'$, again by Def.~\ref{def:lolaDiagnosisAtemp}.
\end{proofof}

\begin{proofof}{Theorem~\ref{thm:multiAggr}}
  (i) Since every $\Delta\in\allDiags$ is by Lemma~\ref{lem:multiIsSingle} a \instDiag{0} for all $t'$ with $t'\in\kpast$, which $\allDiags'$ contains all aggregations from, i.e., every $\Delta'\in\allDiags'$ is also a \instDiag{0} for every $t'$, we have $\Delta\in\allDiags'$. 
  
  (ii) The $\subseteq$ direction follows from (i).
  Suppose for the other direction $\Delta\in\allDiags'$.
  By definition of $\aggr$, this implies $\Delta$ to be a \instDiag{0} for all $t'$, and, by Def.~\ref{def:lolaDiagnosisAtemp}, $C_{t'}:=\{c^{t'}\mid c\in\Delta\}\cup \{\neg c^{t'}\mid c\in\Delta\}\cup\constraintsT{t'}$ to be consistent for those $t'$.
  Because $\varphi$ is atemporal, each instant variable $x^{t'}$ does not have any relation to instant variables $y^{t''}$ with $t''\neq t'$.
  This means that any set of constraints involving instant variables over multiple instants can be split into pairwise disjoint sets of constraints involving only instant variables of a single instant.
  Thus, the consistency of every $C_{t'}$ implies the consistency of $\bigcup_{t'=t-k}^t C_{t'}=\{c^{t'}\mid c\in\Delta\land t'\in\kpast\}\cup\{\neg c^{t'}\mid c\in\comps\setminus\Delta\land t'\in\kpast\}\cup\constraints$ which means, by Definition~\ref{def:lolaDiagnosisAtemp}, that $\Delta\in\allDiags$.

  (iii) Let $\Delta\in\minDiags$.
  To contradict the proposition, there must not exist a $\Delta'\in\minDiags'$ with $\Delta'\subseteq\Delta$, i.e., for all $\Delta'\in\minDiags'$ it would hold that $\Delta\subset\Delta'$.
  However, since by (i) every \instDiag{k} is also in $\aggr(\diags_{t-k},\dotsc,\diags_t)$, i.e., $\Delta\in\allDiags'$, this implies the existence of a (non-minimal) aggregated diagnosis that is a subset of every minimal aggregated diagnosis.
\end{proofof}

\begin{proofof}{Theorem~\ref{thm:multiPrecision}}
  (i) By Definition~\ref{def:lolaDiagnosisAtemp}, $\Delta\in\allDiags$ implies that $C:=\{c^{t'}\mid c\in\Delta\land t'\in\kpast\}\cup \{\neg c^{t'}\mid c\in\comps\setminus\Delta\land t'\in\kpast\}\cup\constraints$ is consistent.
  Therefore, any subset $C'\subseteq C$ is also consistent.
  Let now $C':=\{c^{t'}\mid c\in\Delta\land t'\in\kpastK{k'}\}\cup \{\neg c^{t'}\mid c\in\comps\setminus\Delta\land t'\in\kpastK{k'}\}\cup\constraints$ which, again by Def.~\ref{def:lolaDiagnosisAtemp}, implies $\Delta\in\allDiags'$.

  (ii) Let $\Delta\in\minDiags$.
  To contradict the proposition, there must not exist a $\Delta'\in\minDiags'$ with $\Delta'\subseteq\Delta$, i.e., for all $\Delta'\in\minDiags'$ it would hold that $\Delta\subset\Delta'$.
  However, since by (i) every \instDiag{k} is also a \instDiag{k'}, i.e., $\Delta\in\allDiags'$, this implies the existence of a (non-minimal) \instDiag{k'} that is a subset of every minimal \instDiag{k'}.
\end{proofof}

\begin{proofof}{Theorem~\ref{thm:temp}}
  (i) ($\subseteq$) For each $\Delta\in\allDiags$ there must exist a corresponding $\{(c,t')\mid c\in\Delta\land t'\in\kpast\}\in\allDiags'$ since by Definitions~\ref{def:lolaDiagnosisAtemp} and~\ref{def:lolaDiagnosisTemp} both build the same constraints and must therefore both lead to consistency.
  ($\supseteq$) Let us assume there to be a $\Delta\in\allDiags'$ with $\Delta'=\{c\mid (c,t')\in\Delta\}\not\in\allDiags$.
  Then, let $\Delta''=\{(c,t')\mid c\in\Delta'\land t'\in\kpast\}\supseteq\Delta$ and therefore $\Delta''\in\allDiags'$.
  Hence, $\Delta''$ and $\Delta'$ build the same constraints again by Definitions~\ref{def:lolaDiagnosisAtemp} and~\ref{def:lolaDiagnosisTemp}, leading to $\Delta'\in\allDiags$, and thus contradicting the assumption.

  (ii) Given $\Delta\in\minDiags$, $\Delta'=\{(c,t')\mid c\in\Delta\land t'\in\kpast\}$ is by (i) a \temporalDiag{k}.
  While $\Delta'$ is not necessarily minimal, any minimal $\Delta''\subseteq\Delta'$ ($\Delta''\in\minDiags'$) contains the same components as $\Delta'$, only over possibly fewer timestamps.
  It is not possible that $\Delta''$ contains fewer components than $\Delta'$, i.e. $\{c\mid(c,t')\in\Delta''\}\subset\{c\mid(c,t')\in\Delta'\}$, because that would mean that $\{c\mid(c,t')\in\Delta''\}$ is, by (i), a \instDiag{k} for $t$, leading to $\Delta$ not being minimal due to $\{c\mid(c,t')\in\Delta''\}\subset\Delta$, thus contradicting the assumption.
  Therefore, $\Delta=\{c\mid (c,t')\in\Delta''\}\in\tsFree(\minDiags')$.
\end{proofof}

\end{document}